\documentclass[lettersize,journal]{IEEEtran}
\usepackage{amsmath,amssymb,amsfonts}
\usepackage{algorithmic}
\usepackage{algorithm}
\usepackage{array}
\usepackage[caption=false,font=normalsize,labelfont=sf,textfont=sf]{subfig}
\usepackage{textcomp}
\usepackage{stfloats}
\usepackage{url}
\usepackage{verbatim}
\usepackage{graphicx}
\usepackage{cite, balance}
\usepackage{graphicx,color}
\usepackage{fancyhdr}
\fancypagestyle{firstpage}{
    \fancyhf{}
    \fancyhead[C]{\footnotesize
    This work has been submitted to the IEEE for possible publication.
    Copyright may be transferred without notice, after which this version may no longer be accessible.}
    
}

\newcommand{\bl}[1]{\textcolor{black}{#1}}

\begin{document}

\title{Second-Order Fade Statistics in Underwater Optical Wireless Communications in Air-Bubble and Turbidity-Impaired Environments}

\author{Pedro Salcedo-Serrano, Mohammad-Ali Khalighi, Rubén Boluda-Ruiz, José María Garrido-Balsells, Antonio García-Zambrana%
\thanks{{P. Salcedo-Serrano and M.-A. Khalighi are with Aix-Marseille University, CNRS, Centrale M\'ed, Fresnel Institute, Marseille, France (e-mail: psalserr@gmail.com; Ali.Khalighi@fresnel.fr).}}%
\thanks{{P. Salcedo-Serrano, R. Boluda-Ruiz, J.\,M. Garrido-Balsells, and A. Garc\'ia-Zambrana are also with the University Institute of Telecommunication Research (TELMA), University of M\'alaga, CEI Andaluc\'ia Tech., M\'alaga E-29071, Spain (e-mail: psalserr@gmail.com; rbr@uma.es; jmgb@ic.uma.es; agz@ic.uma.es)}.}%
\thanks{{P. Salcedo-Serrano, J.\,M. Garrido-Balsells, and A. Garc\'ia-Zambrana are also with the Wireless Optical Communications Lab., TELMA}.}%
}



\maketitle
\thispagestyle{firstpage}

\begin{abstract}
The propagation of underwater optical signals through air-bubble-impaired channels induces irradiance fluctuations and partial line-of-sight blockages that degrade the performance of underwater optical wireless communication (UOWC) systems. Although the first-order fading statistics of such channels have been extensively investigated, their second-order statistics, namely the level crossing rate (LCR) and average fade duration (AFD), remain rather unexplored. This paper presents a theoretical and  experimental characterization of the LCR and AFD in bubble-impaired UOWC channels under different bubble sizes and water turbidity conditions. Two analytical approaches are developed based on Rice's formula and on the cumulative distribution function (CDF) together with the bivariate CDF of the sampled irradiance, and are validated experimentally. \bl{The results show that water turbidity has a negligible effect on fade statistics under small-bubble conditions. In contrast, for large bubbles, turbid water leads to shorter but more frequent fades than tap water, as particle-induced beam spreading reduces the likelihood of deep fades and prolonged blockage events.} Finally, the derived second-order statistics are applied to evaluate the \bl{packet outage probability (POP)} performance using a two-state Markov channel model, demonstrating that the higher fade rate in the turbid water case results in a larger POP despite the shorter fade durations.
\end{abstract}


\section{Introduction}
\IEEEPARstart{T}{he} rapid growth of emerging applications, such as real-time video transmission for underwater autonomous systems and environmental ocean monitoring, is driving the demand for high-capacity underwater communication networks~\cite{li_marine_2018, jahanbakht_internet_2021}. Among existing communication technologies, acoustic links suffer from extremely limited bandwidth and high latency, while radio-frequency (RF) links experience severe attenuation in seawater due to its high conductivity. 
Magnetic induction systems, on the other hand, offer low data rates and remain relatively immature from a technological standpoint~\cite{stojanovic_underwater_2009, li_survey_2019}. Consequently, underwater optical wireless communications (UOWC) have emerged as a promising solution for supporting bandwidth-intensive and delay-sensitive applications, including remote offshore infrastructure inspection and maintenance, as well as underwater surveillance missions~\cite{zeng_survey_2017}.

However, the performance of UOWC links is impaired by several factors, including absorption and scattering caused by water molecules and dissolved or suspended particles,  particularly, in relatively high-turbidity waters; beam misalignment; oceanic turbulence induced, for instance, by temperature and salinity gradients; beam blockage due to obstacles such as fish schools; and scattering effects caused by air bubbles~\cite{mobley_oceanic_2022, Gabriel-JOCN-2012, cochenour_temporal_2013, Khalighi-CRC-2017, boluda-ruiz_impulse_2020, boluda-ruiz_capacity_2021, Ijeh-JOCN-2022, ata_analysis_2023, thorpe_bubbles_1980, vahaji_numerical_2018,Nennouche-PhJ-2026}.
Here, we focus on the latter impairment and quantify the performance degradation induced by air bubbles through accurate statistical modeling of their effect under various bubble sizes and water turbidity conditions.
\subsection{Literature Review on UOWC Channel Modeling Under Air Bubble Effects}
From an optical propagation perspective, a cluster of air bubbles can be viewed as a collection of randomly moving local water–air interfaces, owing to both the stochastic nature of bubble generation and the irregular bubble motion induced by ocean currents.
Over the last decade, several studies have statistically characterized and analyzed the impact of air bubbles on UOWC system performance. For instance, \cite{chen_effects_2019} experimentally investigated received signal intensity fluctuations under various air bubble densities and sizes, showing that the best-fitting statistical distribution depends on the specific bubble conditions.
In \cite{shin_statistical_2020}, a simulation-based statistical model was developed to characterize air bubble-induced beam obstruction and power loss, and subsequently integrated with a Gamma–Gamma turbulence model.

In \cite{zedini_unified_2019} and \cite{qiu_unified_2023}, mixture distributions, such as Exponential–generalized Gamma and Weibull–generalized Gamma were proposed to capture the combined impact of air bubble obstruction and oceanic turbulence. In \cite{salcedo-serrano_effect_2024}, an experimental statistical characterization of UOWC channels affected by air bubbles in different water types highlighted the key role of water turbidity in mitigating air bubbles-induced light blockage. Specifically, particle-induced scattering was shown to reduce the scintillation index by enabling the collection of scattered photons, thereby alleviating the effective obstruction caused by large-bubbles or clusters of small-bubbles. Expanding upon these findings, \cite{salcedo-serrano_performance_2024} evaluated UOWC system performance within the proposed empirical statistical framework and confirmed that higher turbidity levels can enhance the overall link bit-error-rate (BER) and outage probability performance by reducing bubble-induced light blockage. \bl{In \cite{kou_investigation_2025}, a combined experimental and theoretical analysis considered} a UOWC channel under the joint effects of air bubbles, turbulence, and suspended particles, leading to the development of a composite channel model capturing their combined impact. The results revealed that turbulence-induced scintillation and path distortion are further exacerbated by air bubbles and particle-induced scattering. \bl{More recently, \cite{Nennouche-PhJ-2026} characterized the water-to-air optical channel by incorporating the effects of wind-generated sea-surface bubbles. Using a Monte Carlo ray-tracing algorithm, the authors evaluated the bubbles' impact on optical propagation based on the Hall-Novarini model and Mie scattering theory.}


\subsection{Need for Second-Order Channel Statistics}
As outlined in the previous subsection, extensive statistical analyses of UOWC channels affected by air bubbles have been conducted under a wide range of conditions, primarily relying on first order statistics, such as probability density functions (PDFs) and outage probability. However, significant gaps remain in the characterization of second-order statistics, notably the level crossing rate (LCR) and average fade duration (AFD), which provide critical insight into the temporal dynamics of fading channels. These metrics are particularly important for the design and optimization of adaptive communication systems, as they directly impact burst error performance and influence the selection of key system parameters, such as the symbol rate, packet and interleaver size~\cite{simon_digital_2005}.

To the best of authors' knowledge, in the UOWC literature, only \cite{salcedo-serrano_effect_2024} briefly introduced AFD as a performance metric based on experimental measurements, yet without providing a theoretical framework. This limited body of work contrasts sharply with the maturity of second-order statistical analysis in atmospheric free-space optical (FSO) channels, where the LCR and AFD have been extensively investigated~\cite{vetelino_fade_2007, jurado-navas_fade_2017, issaid_level_2019, le_design_2019, stefanovic_second_2021, le_level_2021}. Nevertheless, these results cannot be directly extended to UOWC systems, as the physical mechanisms governing irradiance fluctuations induced by air bubbles differ fundamentally from those associated with atmospheric turbulence. Consequently, a rigorous characterization of the LCR and AFD in UOWC channels is essential for understanding the interplay between water turbidity and air bubble dynamics and its impact on system performance.
\subsection{Paper Contributions}
This paper presents an experimental and theoretical analysis of the LCR and AFD in UOWC systems under different bubble sizes and water turbidity conditions. Experimental results are compared with theoretical predictions based on the generalized Gamma and mixture generalized Gamma distributions, for small- and large-bubbles-induced fluctuations, respectively. This work extends the statistical characterization and performance analysis reported in \cite{salcedo-serrano_effect_2024, salcedo-serrano_performance_2024}, by moving beyond first-order channel fading statistics to second-order statistics, namely the LCR and AFD, thereby characterizing the temporal dynamics of bubble-induced fading and the impact of particle-induced scattering. Furthermore, as a direct application of the derived second-order statistics, this work develops analytical expressions for the \bl{packet outage probability} for the considered UOWC scenarios. By accurately capturing the stochastic behavior of air-bubble-induced fading in waters with varying turbidity levels, the presented study provides valuable insights into the temporal dynamics of bubble-induced fading and establishes a framework for the design and optimization of error-control protocols in challenging underwater environments.

In summary, the key contributions of this paper are as follows:
\begin{itemize}
    \item Experimental and theoretical characterization of the LCR and AFD in bubble-impaired UOWC channels under different bubble sizes and turbidity conditions;
    \item Derivation and experimental validation of analytical expressions for the LCR and AFD based on the generalized Gamma and mixture generalized Gamma fading models;
    \item Development of a \bl{packet outage probability} analysis based on a two-state Markov channel model, demonstrating the importance of second-order statistics for packet-level performance evaluation.
\end{itemize}

The remainder of this paper is organized as follows. Section~\ref{sec:UOWCmodel} introduces the UOWC channel model for small- and large-bubble-induced fading. Section~\ref{sec:LCR_AFD} presents two analytical approaches for evaluating the LCR and AFD. Section~\ref{sec:UOWCcharacterization} describes the experimental setup and the statistical characterization of the considered channels. Section~\ref{sec:results} validates the proposed LCR and AFD expressions against experimental results and evaluates the \bl{packet outage probability} 
using a two-state Markov channel model. Finally, Section~\ref{sec:conclusions} concludes the paper and discusses future research directions.

\section{UOWC Channel Model}
\label{sec:UOWCmodel}
In a recent work~\cite{salcedo-serrano_effect_2024}, we proposed an UOWC channel model that distinguishes between \textit{small} and \textit{large} air bubbles, defined relative to the transmitted beam width, as they give rise to distinct fading mechanisms.
Small air bubbles induce small-scale irradiance fluctuations and occasional line-of-sight (LoS) blockage due to bubble clusters. These effects are modeled using a generalized Gamma distribution, given by
\begin{equation}
      f_{I_{S}}(I) = \frac{p}{a^{d}\Gamma(d/p)} \ I^{d-1} \ e^{-(I/a)^{p}},
      \label{eq:GammaGen}
\end{equation}
where $I$ is the normalized received irradiance, $\Gamma(\cdot)$ is the Gamma function, and $a$, $d$, and $p$ are the positive scale and shape parameters of the distribution. The corresponding cumulative distribution function (CDF) is given by \cite{salcedo-serrano_performance_2024}:
\begin{equation}
    F_{I_{S}}(I) = 1 - \frac{\Gamma(\frac{d}{p}, (I/a)^p)}{\Gamma\left(\frac{d}{p}\right)}.
    \label{eq:GG_CDF_upper}
\end{equation}
In contrast, large air bubbles cause frequent LoS blockage events due to their size, \bl{an effect hereafter denoted by $I_B$}, while also inducing small-scale irradiance fluctuations, \bl{an effect hereafter denoted by $I_F$,} through the generation of smaller bubbles resulting from bubble fragmentation, a process driven by natural instabilities or turbulence generated by external propellers.
This dual behavior motivates the use of a mixture model composed of two generalized Gamma distributions, as follows
\begin{equation}
\begin{split}
     f_{I_L}(\bl{I_B, I_F}) & = \,   W   \frac{p_1}{a_1^{d_1}\,\Gamma(d_1/p_1)} \ \bl{I_B}^{d_1-1_1} \ e^{(-\bl{I_B}/a_1)^{p_1}} + \\
     &(1- W) \frac{p_2}{a_2^{d_2}\, \Gamma(d_2/p_2)} \ \bl{I_F}^{d_2-1} \ e^{-(\bl{I_F}/a_2)^{p_2}},
      \label{eq:MixGammaGen}
\end{split}
\end{equation}
where $W \in [0, 1]$ is the mixture weight controlling the relative contribution of each component. As described in \cite{salcedo-serrano_effect_2024}, the first term, \bl{modeling $I_B$ behavior}, captures blockage-dominated fading, whereas the second, \bl{modeling $I_F$ behavior}, accounts for fading dominated by small-scale optical power fluctuations. The corresponding CDF is given by \cite{salcedo-serrano_performance_2024}
\begin{equation}
F_{I_{L}}\left(\bl{I_B}, \bl{I_F}\right)= W \, F_{I_{S}}(\bl{I_B}) + (1-W) \, F_{I_{S}}(\bl{I_F}).
\label{eq:CDF_large}
\end{equation}
\section{Second-Order Statistics: LCR and AFD}
\label{sec:LCR_AFD}
As outlined in the introduction, second-order statistics characterize the temporal evolution of fading beyond its instantaneous amplitude distribution. Here, to investigate UOWC channels affected by air bubbles, we consider the metrics of LCR and AFD.
\bl{The LCR, denoted here by $N_I(I_{\text{th}})$\bl{,} is defined as the average rate at which the received optical signal crosses a certain threshold, $I_{\text{th}}$, in the positive direction. Also, the AFD, denoted by $A_{I}(I_{\text{th}})$, is defined as the average time interval during which the received signal remains below a threshold $I_{\text{th}}$:
\begin{equation}
    A_{I}(I_{\text{th}}) = \frac{F_{I}(I_{\text{th}})}{N_I(I_{\text{th}})},
    \label{eq:AFD}
\end{equation}
where $F_{I}$ is the CDF of the received irradiance.}

\bl{In this work, we investigate two approaches for the estimation of LCR and AFD and validate the derived analytical results through experimental measurements obtained from a UOWC link under various air bubble scenarios and water turbidity conditions.}

\bl{\subsection{Approach 1 (A1): Rice model}
In this rather conventional approach, LCR is calculated using the Rice formula, given by \cite{rice_statistical_1948}
\begin{equation}
    N_I(I_{\text{th}}) = \int_{0}^{+\infty} \dot{I}\, f_{I,\dot{I}}(I_{\text{th}}, \dot{I}) \, d\dot{I},
    \label{eq:LCR_riceApproach}
\end{equation}
where $f_{I,\dot{I}}(I_{th}, \dot{I})$ denotes the joint PDF of the received irradiance $I$ and its time derivative $\dot{I}$. 
This approach has been widely adopted 
in second-order statistical analysis of FSO channels based on the statistics of atmospheric turbulence and its time derivative~\cite{vetelino_fade_2007, jurado-navas_fade_2017, issaid_level_2019, le_design_2019, stefanovic_second_2021, le_level_2021}, and a closed-form expression has been derived for $f_{I,\dot{I}}(I_{th}, \dot{I})$ relying on the specific spatial and temporal statistics of atmospheric turbulence~\cite[Eq. (11.38)]{andrews_laser_2005}.}

\bl{However, deriving the corresponding joint PDF analytically is challenging for the case of air bubble-induced fading in UOWC channels. In fact, to the best of the authors' knowledge, the statistical characterization of air bubble-induced fading in UOWC links has been restricted to the amplitude statistics of the received irradiance, as no theoretical framework currently exists that describes its temporal spectral behavior. Consequently, the LCR approximation derived here relies solely on irradiance amplitude measurements, which represents the most tractable and comprehensive statistical description currently available in the literature.}

Motivated by this, \bl{here,} the LCR approximation is derived under two standard assumptions widely used in wireless fading analysis: (i) the received irradiance $I$ and its time derivative $\dot{I}$ are statistically independent \cite{olutayo_level_2017, hadzi-velkov_level_2007, cotton_second-order_2016}, and (ii) $\dot{I}$ follows a zero-mean Gaussian distribution \cite{vetelino_fade_2007, jurado-navas_fade_2017, issaid_level_2019, le_level_2021, andrews_laser_2005}. Accordingly, we have
\begin{equation}
    f_{I,\dot{I}}(I,\dot{I}) = f_I(I) f_{\dot{I}}(\dot{I}),
    \label{eq:jointPDF}
\end{equation}
\begin{equation}
    f_{\dot{I}}(\dot{I}) = \frac{1}{\sqrt{2\pi \sigma^2_{\dot{I}}}} \exp\left(-\frac{\dot{I}^2}{2\sigma^2_{\dot{I}}}\right),
    \label{eq:dotI}
\end{equation}
where $\sigma^2_{\dot{I}}$ is the variance of the time derivative of the received irradiance. Then, from (\ref{eq:LCR_riceApproach}), the LCR can be approximated as
\begin{equation}
    N_{I, A1}(I_{\text{th}}) \approx f_{I}(I_{{\text{th}}}) \int_{0}^{+\infty} \dot{I}\, f_{\dot{I}}(\dot{I}) \, d\dot{I},
    \label{eq:LCR_approx}
\end{equation}
\bl{where $N_{I, A1}(I_{\text{th}})$ denotes the LCR based on Approach~1.} Substituting (\ref{eq:dotI}) into (\ref{eq:LCR_approx}), 
\begin{equation}
\begin{split}
    \displaystyle\int_{0}^{+\infty} \dot{I}\, f_{\dot{I}}(\dot{I}) \, d\dot{I} = &
    \frac{1}{\sqrt{2\pi\sigma^2_{\dot{I}}}} \int_{0}^{+\infty} \dot{I} \
     e^{\big(-\frac{\dot{I}^2}{2\sigma^2_{\dot{I}}}\big)} d\dot{I}=\frac{\sigma_{\dot{I}}}{\sqrt{2\pi}}.  
    \label{eq:integral}
\end{split}
\end{equation}
Therefore,
\begin{equation}
    N_{I, A1}(I_{{\text{th}}}) \approx f_{I}(I_{{\text{th}}}) \frac{\sigma_{\dot{I}}}{\sqrt{2\pi}}.
    \label{eq:LCR_final}
\end{equation}

In terrestrial FSO links, $\sigma^2_{\dot{I}}$ is related to the quasi-frequency $\nu_{0}$, which quantifies the spectral width of the normalized irradiance power spectrum~\mbox{\cite[Eq. (11.38)]{andrews_laser_2005}}.
However, the normalized irradiance power spectrum of bubble-impaired UOWC channels remains unknown. Therefore,  we estimate $\sigma^2_{\dot{I}}$ directly from the empirical autocorrelation of the irradiance fluctuation, $B_{I}(\tau)$, as follows \mbox{\cite[Eq. (6.6-20)]{beckmann_probability_1967}}
\begin{equation}
    \sigma^2_{\dot{I}} = B_{\dot{I}}(0) = - B''_{I}(0),
    \label{eq:sigma_dotI}
\end{equation}
where $B''_{I}(\tau)$ represents the second-order time derivative of $B_{I}(\tau)$. This relationship holds under the assumption that $\mathbb{E}[\dot{I}] = 0$ ($\mathbb{E}[.]$ denoting the expected value), which is valid for a stationary process. Using the Wiener-Khinchin theorem together with the relationship between the power spectral density (PSD) of a stationary random process and its derivative \mbox{\cite[Eq. (6.6-28)]{beckmann_probability_1967}}, $B''_{I}(0)$ can be obtained as \mbox{\cite[Eq. (6.6-29)]{beckmann_probability_1967}}
\begin{equation}
    \label{eq:PSD}
    - B''_{I}(0) = \frac{1}{2\pi}\int \omega^2 S_{I}(\omega) d\omega,
\end{equation}
where $\omega$ is the angular frequency and $S_{I}(\omega)$ is the PSD of the empirical measurement. 

\bl{Given $N_{I, A1}(I_{{\text{th}}})$, the AFD is then calculated by substituting (\ref{eq:LCR_final}) into (\ref{eq:AFD}) as follows
\begin{equation}
    A_{I, A1}(I_{\text{th}}) \approx \sqrt{2\pi} \frac{F_{I}(I_{\text{th}})}{f_{I}(I_{{\text{th}}}) \sigma_{\dot{I}}}.
    \label{eq:AFD_Approach1}
\end{equation}}

\subsection{Approach 2 (A2): Sampled Fading Channels}
\bl{An alternative method for estimating the LCR of sampled fading processes was proposed in~\cite{lopez-martinez_higher_2012}.
This approach yields a closed-form LCR expression that inherently accounts for the discrete-time nature of sampled fading measurements~\cite{lopez-martinez_bivariate_2013}. Specifically, the LCR is expressed in terms of the CDF and bivariate CDF of the sampled envelope, evaluated at the threshold $I_{\text{th}}$, as follows:}
\begin{equation}
    N_{I, A2}(I_{\text{th}})=\frac{F_I(I_{\text{th}})-F_{I_1,I_2}(I_{\text{th}},I_{\text{th}})}{T_S},
    \label{eq:LCR_bivariateApproach}
\end{equation}
where $T_S$ is the sampling period, $I_1 \triangleq I(t)$, $I_2 \triangleq I(t+T_S)$, and $F_{I_1,I_2}$ denotes the bivariate CDF \cite{lopez-martinez_higher_2012}. 
\bl{Given (\ref{eq:AFD}), the AFD is expressed as}
\begin{equation}
    A_{I, A2}(I_{\text{th}}) = T_S \left(1 - \frac{F_{I_1,I_2}(I_{\text{th}}, I_{\text{th}})}{F_I(I_{\text{th}})} \right)^{-1}.
\end{equation}
\section{Empirical UOWC Channel Characterization}
\label{sec:UOWCcharacterization}
\subsection{Experimental Setup}
\begin{figure*}[!t]
\centering
\begin{minipage}{2\columnwidth}
    \centering
    \includegraphics[width=0.8\linewidth]{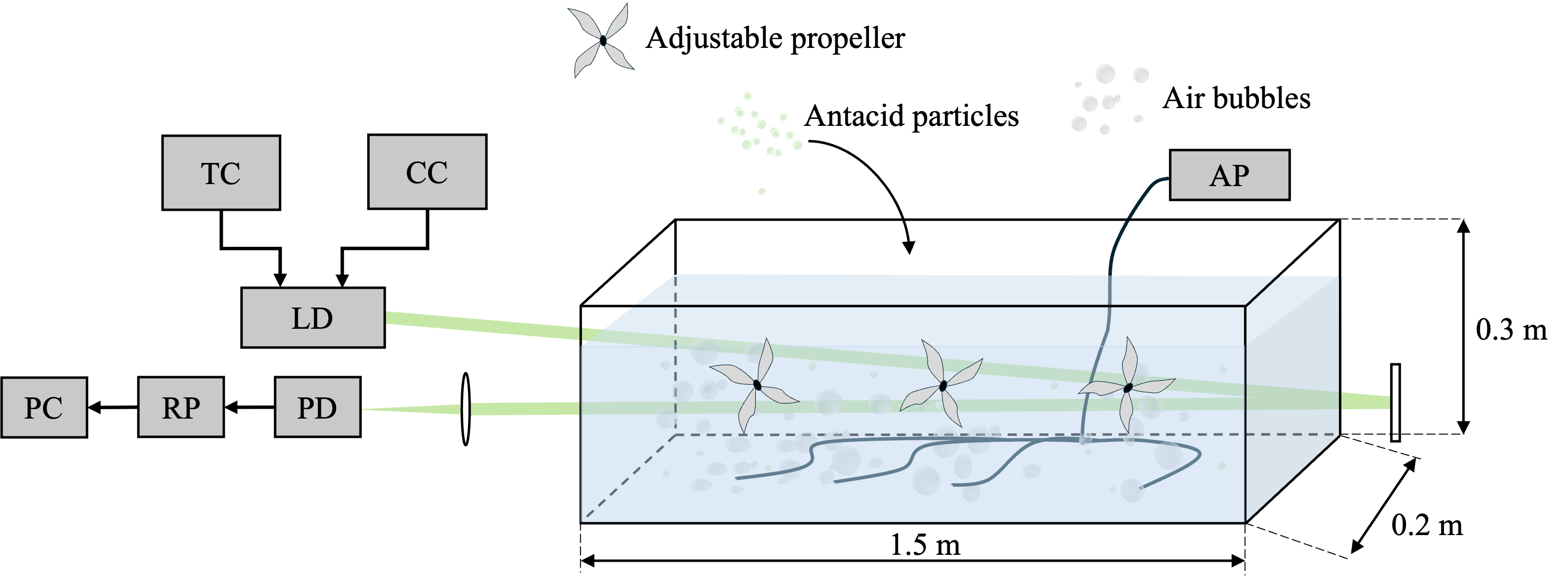} 
    
    (a)
\end{minipage}
\hfill
\vspace{0.3cm}
\begin{minipage}{2\columnwidth}
    \centering
    \includegraphics[width=0.9\linewidth]{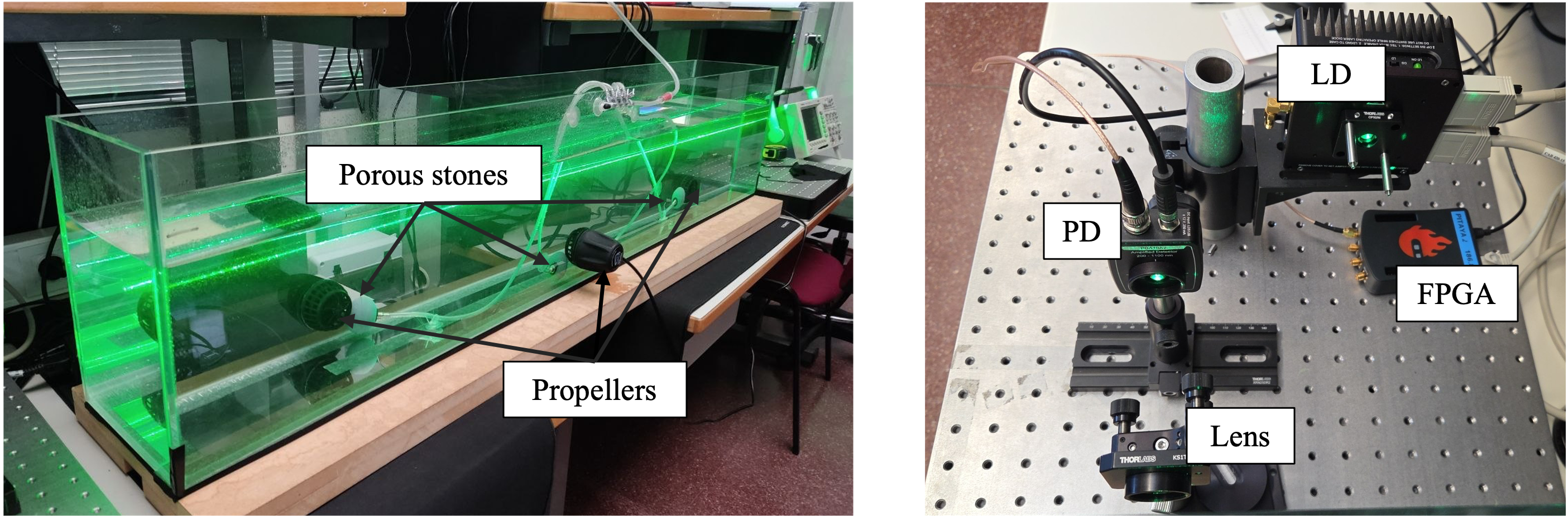}
    
    (b)
\end{minipage}
\caption{Experimental setup for UOWC channel measurements. (a) Schematic of the setup; PD: photodiode, LD: laser diode, TC: temperature controller, CC: current controller, AP: air pump, PC: personal computer, RP: Red Pitaya STEMLab. (b) Snapshot of the setup, showing the water tank (left), and the transmitter/receiver components (right).}
\label{fig:setup}
\end{figure*}

The block diagram and a photograph of the experimental setup are shown in Fig.~\ref{fig:setup}. The setup is based on that used in our recent work~\cite{salcedo-serrano_effect_2024}, with two key modifications introduced to improve  measurement accuracy and better emulate natural underwater scattering conditions.  First, water turbidity is reproduced using a mixture of \mbox{Mg(OH)$_{2}$} and \mbox{Al(OH)$_{3}$} antacids, whose volume scattering function closely matches that of natural water constituents~\cite{mullen_investigation_2011}, unlike the commercial Maalox suspension used in~\cite{salcedo-serrano_effect_2024}. Second, a higher-power laser diode ($115$~mW output power) is employed to improve the receiver signal-to-noise ratio (SNR), thereby enabling more accurate statistical measurements.\footnote{The transmitter uses a TO-can green laser diode (Thorlabs L520A2) operating at $520$~nm with an output power of $115$~mW. The LD is mounted on a temperature-controlled mount (Thorlabs LDM56/M) maintained at 20°C by a temperature controller (Thorlabs TED200C) and an integrated temperature transducer (Analog Devices AD592). \bl{The laser output power in \cite{salcedo-serrano_effect_2024} was 15~mW.}}

After propagating through the UOWC channel, the optical signal is collected by a silicon \bl{PIN} photodiode receiver (Thorlabs PDA10A2) equipped with a $25.4$~mm diameter, $125$~mm focal-length lens (Thorlabs LA1986-A). The photodiode output is digitized using a Red Pitaya STEMlab at a sampling frequency of $f_s = 2.5 \text{ MHz}$, yielding $6 \times 10^8$ samples \bl{(240 seconds)} that are transferred via IP to an external computer for offline processing. A low-pass filter (LPF) is applied to the sampled data in order to restrict the statistical analysis to the bandwidth of the optical fluctuations, thereby reducing the contribution of external noise sources. Since the channel coherence time associated with air bubbles is typically on the order of a few milliseconds~\cite{salcedo-serrano_effect_2024}, the LPF cut-off frequency $f_c$ should be on the order of several kHz. Here, we select \mbox{$f_c = 5$~kHz} to preserve fine spectral components that, despite their low energy, have a non-negligible impact on LCR and AFD. 

The underwater channel emulator consists of a $1.5 \times 0.2 \times 0.3$ m$^3$ glass tank filled with $55$~L of water. Ocean currents are emulated by three brushless DC motors driving propellers with adjustable speed and configurable flow pattern. A broadband dielectric mirror (Thorlabs BBDSQ-E02) is placed at one end of the tank to fold the optical path, extending the UOWC link length to $3$~m. Air bubbles are generated by an external air pump connected to four porous stones distributed along the tank, providing a constant airflow rate of $16$~L/min. The bubble size regime is controlled by the porous stones: small-bubbles are generated with the stones in place, whereas large-bubbles are produced after their removal, as described in~\cite{salcedo-serrano_effect_2024}. The porous stones were arranged to produce comparable path loss and scintillation index values for both bubble-size scenarios. Two water turbidity conditions are considered: a clear-water baseline using untreated tap water and a turbid-water condition obtained by dissolving 100~mg each of Mg(OH)$_{2}$ and Al(OH)$_{3}$
antacids (total of $200$~mg) in the tank.

\subsection{Path Loss and Scintillation Index}
We investigate four transmission scenarios corresponding to different combinations of water turbidity and air bubble conditions. The mean path loss and normalized irradiance variance (i.e., the scintillation index) measured for each scenario are reported in Table \ref{tab:pathloss}. As can be observed, the measured path loss differs only slightly between the small- and large-bubble scenarios in tap water (less than $1$~dB). After the introduction of the antacid, the received optical power decreases significantly due to particle-induced absorption and scattering, increasing the path loss to approximately $-34$~dB. In fact, as particle-induced scattering dominates the channel attenuation, the path loss is nearly the same for the small- and large-bubble scenarios. On the other hand, a marked difference is observed between the two bubble regimes when turbidity is introduced regarding the scintillation index. For large-bubbles, the scintillation index decreases from $0.168$ to $0.137$ after adding the antacid, in agreement with the findings of~\cite{salcedo-serrano_effect_2024}. In contrast, the scintillation index remains virtually unchanged in the small-bubble scenario. This behavior can be explained by the fact that small-bubbles rarely form clusters large enough to produce significant LoS blockage. Consequently, since the probability of LoS blockage is already negligible, the additional particle-induced scattering does not produce any measurable reduction in irradiance fluctuations.

\begin{table}[!t]
    \centering
    \caption{Mean received optical power $\overline{P_{\mathrm{rx}}}$, estimated path loss $L$, and scintillation index $\sigma^2_{I}$, for small- and large-bubble scenarios under clear (tap) and turbid water conditions.}
    \begin{tabular}{llccc}
        \hline
        \textbf{Scenario} & \textbf{Antacid level} &
        $\overline{P_{\mathrm{rx}}}$ (mW) &
        $L$ (dB) & $\sigma^2_{I}$\\
        \hline
        \hline
        Small-bubbles & 0 mg   
            & 8.99 & -11.11 
            & 0.164\\
        Small-bubbles & 200 mg 
            & 47.94 $\times 10^{-3}$ & -33.84
            & 0.165\\
        \hline
        Large-bubbles & 0 mg   
            & 10.56 & -10.41 
            & 0.168 \\
        Large-bubbles & 200 mg 
            & 47.93 $\times 10^{-3}$ & -33.84 
            & 0.137 \\
        \hline
    \end{tabular}
    \label{tab:pathloss}
\end{table}
\subsection{PDF and CDF Characterization}
Using the two approaches of A1 and A2 presented in Section~\ref{sec:LCR_AFD}, we estimate the empirical histogram and CDF of the normalized received optical power, with $\mathbb{E}[I] = 1$. This normalization allows a fair comparison of the fading second order statistics between tap water and turbid water scenarios. 

In the case of the Rice Approach (A1), the distribution parameters \bl{of $f_{I}(\cdot)$} are estimated by minimizing the mean squared error (MSE) with respect to the empirical PDF, given by \cite{fernandez_how_2025}
\begin{equation}
    \label{eq:error_MSE_Approach1}
    \epsilon_{\mathrm{MSE, A_1}}\left( \boldsymbol{\lambda}\right) 
    \triangleq \frac{1}{K} \sum_{k=1}^{K} 
    \left(\hat{f}_I(x_k) - f_I(x_k; \boldsymbol{\lambda})\right)^2,
\end{equation}
where $\hat{f}_I(\cdot)$ is the empirical PDF, $f_I(\cdot)$ is the fitted PDF, $K$ is the number of points at which the histogram is estimated, $x_k$ denotes the abscissa of the $k$-th histogram point, and $\boldsymbol{\lambda}$ is the set of estimated distribution parameters. Specifically, $\boldsymbol{\lambda} = \{a, d, p\}$ for the small-bubble scenario, and \mbox{$\boldsymbol{\lambda} = \{W, a_1, d_1, p_1, a_2, d_2, p_2\}$} for the large-bubble scenario, see (\ref{eq:GammaGen}) and (\ref{eq:MixGammaGen}). The resulting fitted parameters and the corresponding MSE values are listed in Table \ref{tab:parameters_PDF_Approach1}. 

\begin{table*}[ht]
    \centering
    \caption{Estimated parameters of the generalized Gamma distribution and mixture model \bl{for Approach 1 in} small and large-bubbles scenarios, respectively, under tap water (0 mg) and turbid water (200 mg) conditions.}
    \label{tab:parameters_PDF_Approach1}
    \renewcommand{\arraystretch}{1.3}
    \begin{tabular}{llcccccc}
        \hline
        \textbf{Scenario} & \textbf{Antacid level} & $W$ & $d$ ($d_1$/$d_2$) & 
        $a$ ($a_1$/$a_2$) & $p$ ($p_1$/$p_2$) & 
        $\epsilon_{\mathrm{MSE}, A_1}$ [\bl{Eq. (\ref{eq:error_MSE_Approach1})}] \\
        \hline
        \hline
        Small-bubbles & 0 mg 
            & --- 
            & $2.2504$ 
            & $1.3553$ 
            & $3.8333$ 
            & $3.2975 \times 10^{-4}$ \\
        Small-bubbles & 200 mg
            & --- 
            & $2.8119$ 
            & $1.0772$ 
            & $2.5867$ 
            & $3.1279 \times 10^{-3}$ \\
        \hline
        Large-bubbles & 0 mg
            & $0.1177$ 
            & $2.8823$ / $3.6184$ 
            & $1.4130 \times 10^{-6}$ / $1.4074$ 
            & $0.2270$ / $7.8569$ 
            & $2.1335 \times 10^{-3}$ \\
        Large-bubbles & 200 mg 
            & $0.0901$ 
            & $30.9787$ / $3.9165$ 
            & $2.6335 \times 10^{-18}$ / $1.3479$ 
            & $0.1401$ / $8.8568$ 
            & $5.8969 \times 10^{-3}$ \\
        \hline
    \end{tabular}
\end{table*}

In the case of Approach 2, the parameters of the fitted distribution \bl{$F_I(\cdot)$ and $F_{I_1,I_2}(\cdot, \cdot)$} are obtained by minimizing the MSE with respect to the empirical CDF, given by
\begin{equation}
    \label{eq:error_MSE_Approach2}
    \epsilon_{\mathrm{MSE, A_2}}\left(\boldsymbol{\lambda}\right) 
    \triangleq \frac{1}{K} \sum_{k=1}^{K} 
    \left(\hat{F}_I(x_k) - F_I(x_k; \boldsymbol{\lambda})\right)^2.
\end{equation}

It is worth noting that the full characterization of $F_{I_1,I_2}$ is not necessary, since the joint CDF evaluated at equal arguments, $I_\text{th}$, reduces to the CDF of $I_M \triangleq \text{max}\{I_1, I_2\}$~\cite{lopez-martinez_higher_2012}. The realization of $I_M$ is directly obtained from the experimental measurements by forming consecutive sample pairs $(I_1[n], I_2[n]) = (r[n], r[n+1])$, and retaining, for each pair, the larger of the two values, i.e., $I_M[n]=\max (I_1[n], I_2[n])$. Then, the parameters of $F_{I_M}(\cdot; \boldsymbol{\lambda})$ are obtained in a similar way by fitting a generalized Gamma distribution to the empirical CDF of $I_{M}$. \bl{It should be noted that $I_M$ is fitted using a generalized Gamma distribution for small air bubbles, as described in (\ref{eq:GammaGen}), and a mixture of two generalized Gamma distributions for large air bubbles, as described in (\ref{eq:MixGammaGen}). The resulting fitted parameters and the corresponding MSE values for both $I$ and $I_M$ are reported in Table \ref{tab:parameters_PDF_Approach2}}.


\begin{table*}[ht]
    \centering
    \caption{\bl{Estimated parameters of the generalized Gamma distribution and mixture model for Approach 2 in small and large-bubbles scenarios, respectively, under tap water (0 mg) and turbid water (200 mg) conditions.}}
    \label{tab:parameters_PDF_Approach2}
    \renewcommand{\arraystretch}{1.3}
    \begin{tabular}{llcccccc}
        \hline
        \textbf{Scenario} & \textbf{Antacid level} & $W$ & $d$ ($d_1$/$d_2$) & 
        $a$ ($a_1$/$a_2$) & $p$ ($p_1$/$p_2$) & 
        $\epsilon_{\mathrm{MSE}, A_2}$ [\bl{Eq. (\ref{eq:error_MSE_Approach2})}] \\ 
        \hline
        \hline
        \multicolumn{7}{c}{$I$}\\
        \hline
        Small-bubbles & 0 mg 
            & --- 
            & $2.2928$ 
            & $1.3326$ 
            & $3.5938$ 
            & $2.8076 \times 10^{-6}$ \\
        Small-bubbles & 200 mg
            & --- 
            & $2.8506$ 
            & $1.0569$ 
            & $2.5115$ 
            & $2.4350 \times 10^{-6}$ \\
        \hline
        Large-bubbles & 0 mg
            & $0.1185$ 
            & $2.9927$ / $3.0733$ 
            & $6.0401 \times 10^{-7}$ / $1.4775$ 
            & $0.2196$ / $12.5673$ 
            & $2.2122 \times 10^{-6}$ \\
        Large-bubbles & 200 mg 
            & $0.0976$ 
            & $10.9144$ / $3.7887$ 
            & $2.1986 \times 10^{-6}$ / $1.3716$ 
            & $0.3162$ / $8.7508$ 
            & $5.1867 \times 10^{-6}$ \\
        \hline
        \hline
        \multicolumn{7}{c}{$I_M$}\\
        \hline
        Small-bubbles & 0 mg 
            & --- 
            & $2.2929$ 
            & $1.3326$ 
            & $3.5938$ 
            & $2.8074 \times 10^{-6}$ \\
        Small-bubbles & 200 mg
            & --- 
            & $2.8507$ 
            & $1.0569$ 
            & $2.5115$ 
            & $2.4349 \times 10^{-6}$ \\
        \hline
        Large-bubbles & 0 mg
            & $0.1185$ 
            & $2.9937$ / $3.0733$ 
            & $6.0005 \times 10^{-7}$ / $1.4775$ 
            & $0.2195$ / $12.5680$ 
            &  $2.2147 \times 10^{-6}$ \\
        Large-bubbles & 200 mg 
            & $0.0976$ 
            & $10.9038$ / $3.7887$ 
            & $2.2488 \times 10^{-6}$ / $1.3717$ 
            & $0.3167$ / $8.7515$ 
            & $5.1890 \times 10^{-6}$ \\
        \hline
        \hline
    \end{tabular}
\end{table*}

\begin{figure}[!t]
\centering
\begin{minipage}{1\columnwidth}
    \centering
    \includegraphics[width=\linewidth]{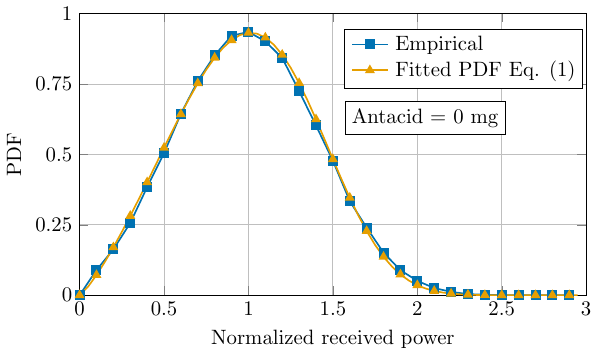}
    (a) 
\end{minipage}
\hfill
\begin{minipage}{1\columnwidth}
    \centering
    \includegraphics[width=\linewidth]{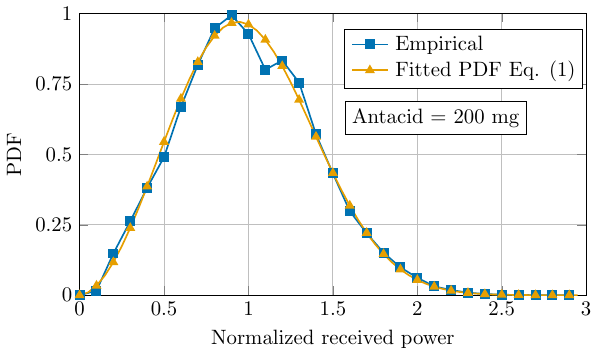}
    (b)
\end{minipage}
\caption{Comparison of the empirical and fitted probability density function of the {\bf small-bubbles} for (a) tap water and (b) turbid water ($200$~mg antacid) conditions.}
\label{fig:PDF_small}
\end{figure}

\begin{figure}[!t]
\centering
\begin{minipage}{1\columnwidth}
    \centering
    \includegraphics[width=\linewidth]{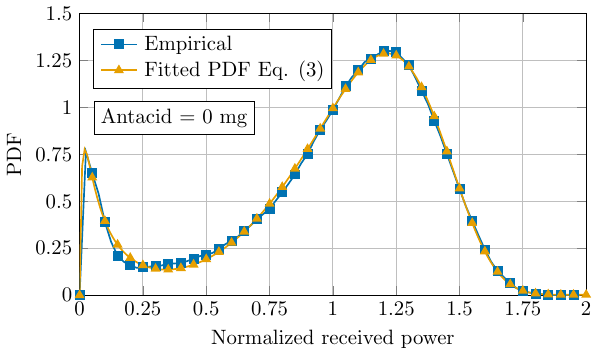}
    (a) 
\end{minipage}
\hfill
\begin{minipage}{1\columnwidth}
    \centering
    \includegraphics[width=\linewidth]{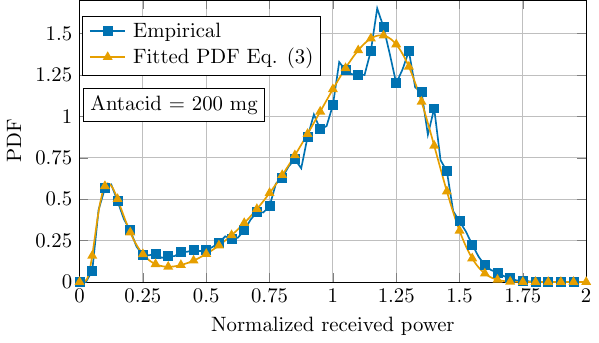}
    (b)
\end{minipage}
\caption{Comparison of the empirical and fitted probability density function of the {\bf large-bubbles} for (a) tap water and (b) turbid water ($200$~mg antacid) conditions.}
\label{fig:PDF_large}
\end{figure}

Figures~\ref{fig:PDF_small} and \ref{fig:PDF_large} show the empirical PDF alongside the fitted distributions \bl{following (\ref{eq:error_MSE_Approach1})} for all considered scenarios, evaluated at the parameter sets $\boldsymbol{\lambda}$ listed in Table \ref{tab:parameters_PDF_Approach1} for the cases of small and large air bubbles, respectively. The fitted PDFs closely match the empirical PDFs in all cases (\bl{as evidenced by the low MSE values reported in Table \ref{tab:parameters_PDF_Approach1}}) confirming the suitability of the proposed models. As expected, the PDF for the small-bubble case in Fig.~\ref{fig:PDF_small} is unimodal and centered around the mean received power, consistent with small-scale irradiance fluctuations and negligible LoS blockage. In contrast, the histogram for the large-bubble scenario in Fig.~\ref{fig:PDF_large} exhibits a clear bimodal distribution, reflecting the two fading mechanisms captured by the proposed mixture model. Here, the lower peak corresponds to blockage-dominated events, in which large-bubbles obstruct the LoS, whereas the higher peak represents irradiance fluctuations around the mean received power, where the channel behaves similarly to the small-bubbles case. Notably, in the turbid water case, the lower peak becomes less pronounced and shifts toward the mean, indicating that turbidity mitigates the severity of blockage events, consistent with the observations reported in \cite{salcedo-serrano_effect_2024}. 

Note that the histogram for the turbid-water case exhibits higher ``noise'' due to the lower received signal-to-noise ratio. However, this does not affect the distribution shape or the quality of the parametric fit. Also, for the large-bubble case in Fig.~\ref{fig:PDF_large}, the fitted scale parameter $a_1$ is extremely small under both water conditions, indicating that blockage events reduce the received irradiance to nearly zero, as represented by the first component of the proposed mixture model.

\begin{figure}[!t]
\centering
\begin{minipage}{1\columnwidth}
    \centering
    \includegraphics[width=\linewidth]{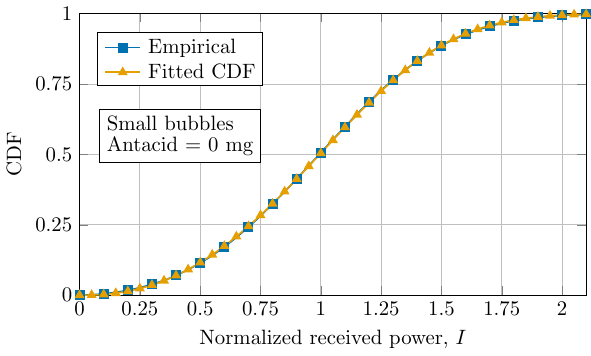}
    (a) 
\end{minipage}
\hfill
\begin{minipage}{1\columnwidth}
    \centering
    \includegraphics[width=\linewidth]{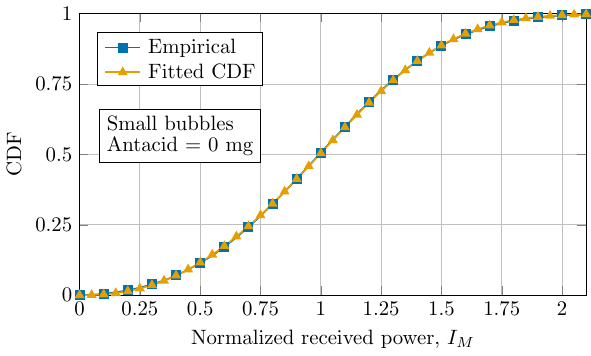}
    (b)
\end{minipage}
\caption{\bl{Comparison of the empirical and fitted cumulative density functions under small-bubble conditions for tap water ($0$~mg of antacid): (a) $I$, and (b) $I_M$.}}
\label{fig:CDF_Approach2_small_0mg}
\end{figure}

\bl{Similarly, Fig.~\ref{fig:CDF_Approach2_small_0mg} 
compares the empirical CDFs with the distributions fitted using (\ref{eq:error_MSE_Approach2}) for the small-bubble scenario under tap-water conditions, based on the parameter set $\boldsymbol{\lambda}$ reported in Table \ref{tab:parameters_PDF_Approach2}. 
The fitted distributions closely match the empirical CDFs for both $I$ and $I_M$ (\bl{as evidenced by the low MSE values reported in Table \ref{tab:parameters_PDF_Approach2}}). For conciseness, only this representative case is shown, as the remaining scenarios exhibit comparable fitting accuracy and provide no additional insight into the goodness of fit. Moreover, although not readily discernible from the figure, the condition $F_{I_M}(\cdot) \leq F_{I}(\cdot)$ holds over the entire range of interest considered in this work, thereby preventing physically meaningless negative LCR values in (\ref{eq:LCR_bivariateApproach}).}

\section{Statistical Characterization of Fading}
\label{sec:results}
In this section, we present numerical results based on the derived PDFs to analyze the LCR and AFD characteristics of UOWC channels in the presence of small and large air bubbles under different levels of water turbidity. To that end, the empirical LCR and AFD are computed from each normalized irradiance measurement after LPF filtering over a wide range of thresholds spanning the dynamic range of the normalized received signal. \bl{Mathematically, the empirical LCR at threshold $I_{\text{th}}$ in the positive-going direction is estimated as
\begin{equation}
    \widehat{N}_{I}(I_{\text{th}}) = \frac{1}{T_{\mathrm{obs}}}
    \sum_{n=1}^{N-1} \mathbf{1}\left\{I[n] \leq I_{\text{th}},\; I[n+1] > I_{\text{th}}\right\},
    \label{eq:LCR_emp}
\end{equation}
where $T_{\mathrm{obs}}$ is the observation period (i.e., the measurement duration), $N$ is the total number of samples, and $\mathbf{1}\{\cdot\}$ denotes the indicator function, which equals unity when the signal is at or below $I_{\text{th}}$ at sample $n$ and exceeds $I_{\text{th}}$ at sample $n+1$, and zero otherwise. To assess the consistency of the experimental results and quantify the variability of the estimated LCR and AFD, the $6 \times 10^8$ samples were partitioned into $120$ contiguous, non-overlapping blocks of $5 \times 10^6$ samples ($2$~s) each, and $\widehat{N}_{I}(I_{\text{th}})$ was estimated for each block. The block duration was chosen to be much longer than the channel coherence time, which is at most on the order of $1$~ms for all considered scenarios \cite{salcedo-serrano_effect_2024}. For each threshold, the standard deviations of the blockwise LCR and AFD estimates were then computed, providing threshold-dependent measures of their statistical variability over the observation period.}

In addition, we compare the empirical results with the analytical predictions obtained from the two approaches derived in (\ref{eq:LCR_final}) and (\ref{eq:LCR_bivariateApproach}).

\subsection{Level Crossing Rate Analysis}
The LCR for small- and large-bubble scenarios is plotted as a function of $I_{\mathrm{th}}$ in Figs.~\ref{fig:LCR_small} and \ref{fig:LCR_large}, respectively, where both tap water and turbid water conditions are considered.\\

\subsubsection{Experimental LCR results}
\bl{The mean experimental LCR across blocks closely agrees with the estimate obtained from the full measurement, supporting the approximate ergodicity of the signal over the observation period. The vertical error bars at each experimental point represent the standard deviation across blocks. As observed, the variability increases markedly at low threshold levels, where crossing events become rare \bl{(as evidenced by the PDFs in Figs.~\ref{fig:PDF_small} and \ref{fig:PDF_large}), }
resulting in greater statistical uncertainty in the LCR estimate.
}

\begin{figure}[!t]
\centering
\begin{minipage}{1\columnwidth}
    \centering
    \includegraphics[width=\linewidth]{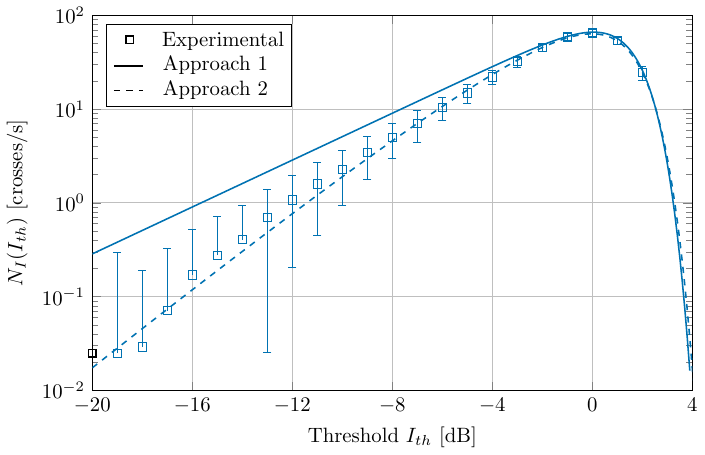}
    (a) 
\end{minipage}
\hfill
\begin{minipage}{1\columnwidth}
    \centering
    \includegraphics[width=\linewidth]{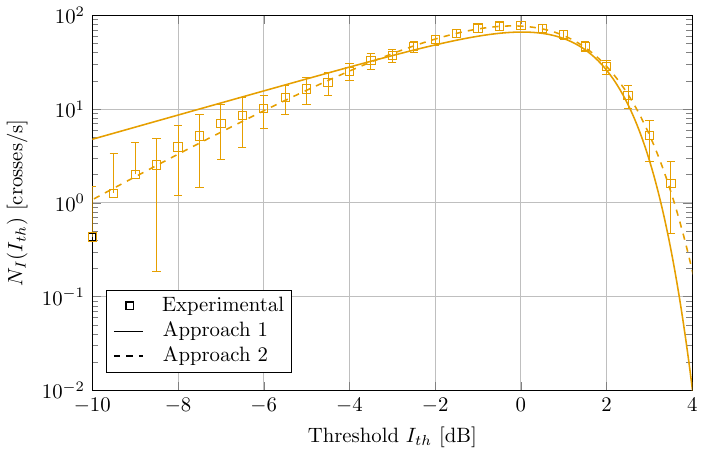}
    (b)
\end{minipage}
\caption{Average LCR versus a threshold $I_{\mathrm{th}}$ for (a) tap water and (b) turbid water ($200$~mg of antacid) in the presence of {\bf small air bubbles}.}
\label{fig:LCR_small}
\end{figure}

Consider first the small-bubble scenario. According to Fig.~\ref{fig:LCR_small}, the analytical LCR closely matches the experimental results for both water conditions, confirming the validity of the proposed models. The observed LCR behavior is also consistent with previous studies reported in the literature \cite{vetelino_fade_2007, jurado-navas_fade_2017, issaid_level_2019, le_design_2019, stefanovic_second_2021, le_level_2021}. The LCR reaches its maximum at approximately \mbox{$I_{\mathrm{th}}= 0 $ dB}, corresponding to the mean normalized irradiance, where threshold crossings occur most frequently. As $I_{\mathrm{th}}$ decreases below $0$~dB, the LCR decreases because deep fades occur less frequently. Likewise, for thresholds above $0$~dB, the LCR also decreases, since the received irradiance rarely exceeds the mean by a large margin. However, it is worth noting that the LCR curve is markedly asymmetric, in contrast to the nearly symmetric behavior typically reported for FSO channels under atmospheric turbulence \cite{issaid_level_2019}. Specifically, the LCR decays more gradually for \mbox{$I_{\mathrm{th}}< 0 $~dB} than for \mbox{$I_{\mathrm{th}}> 0 $~dB}. This asymmetry is attributed to the behavior of small-bubble-induced fading. On one hand, bubble clusters can produce deep fades, increasing the number of crossings at low thresholds. On the other hand, these fades evolve more rapidly than positive irradiance excursions, further increasing the crossing rate. Consequently, small-bubbles generate short, consecutive deep fades, whereas high-irradiance states persist for longer durations in the absence of partial LoS blockage. 

\bl{Additionally, as expected, both water turbidity conditions exhibit similar LCR trends \bl{and comparable values over the common threshold range, with approximately 100 crossings/s at \mbox{$I_{\mathrm{th}} = 0 $~dB}, and 10 crossings/s near \mbox{$I_{\mathrm{th}} = -6 $~dB}}. Since small bubbles primarily induce small-scale irradiance fluctuations with negligible LoS blockage, the additional scattering-induced beam spreading the turbid-water scenario has little impact on the LCR behavior. Some differences arise for \mbox{$I_{\mathrm{th}} < -7.5$~dB}, where lower LCR values are observed in the turbid-water case. However, these differences are likely attributable to the scarcity of crossing events at such low threshold levels, which makes the estimator subject to greater statistical uncertainty, as reflected by the large standard-deviation bars in this region for both  water conditions.}

\begin{figure}[!t]
\centering
\begin{minipage}{1\columnwidth}
    \centering
    \includegraphics[width=\linewidth]{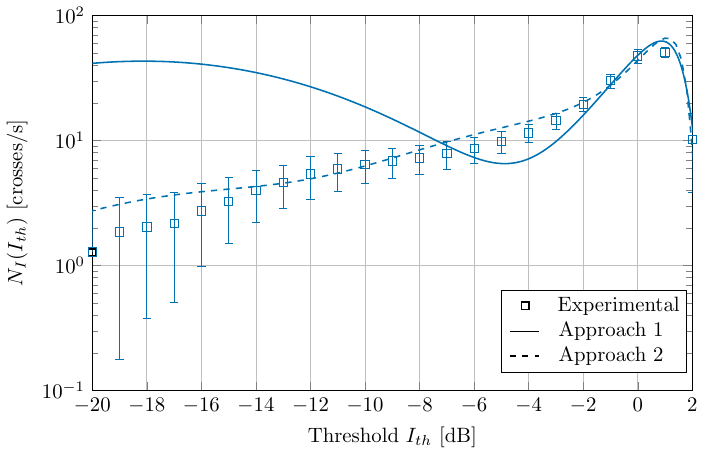}
    (a)
\end{minipage}
\hfill
\begin{minipage}{1\columnwidth}
    \centering
    \includegraphics[width=\linewidth]{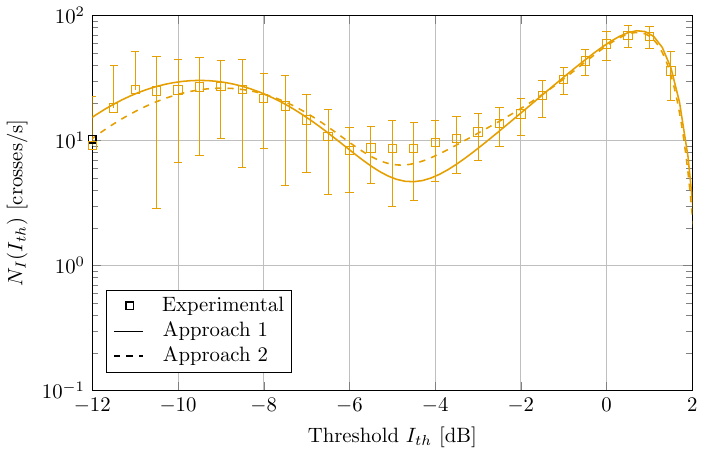}
    (b) 
\end{minipage}
\caption{Average LCR versus a threshold $I_{\mathrm{th}}$ for (a) tap water and (b) turbid water ($200$~mg of antacid) in the presence of \textbf{large air bubbles}.}
\label{fig:LCR_large}
\end{figure}

Now consider the large-bubble scenario. As seen in Fig.~\ref{fig:LCR_large}, \bl{the LCR here exhibits a markedly different behavior depending on the water condition. For the tap-water case, the experimental LCR displays a single well-defined peak near $0$~dB, followed by a gradual monotonic decay toward lower thresholds. No secondary peak is discernible in the experimental data in this region, where the large standard-deviation bars also indicate increased statistical uncertainty at low thresholds. In contrast, under turbid-water conditions, the experimental LCR exhibits a clear bimodal profile, with a primary peak near $1$~dB and a secondary peak near $-9$~dB, consistent with the PDF behavior observed in Fig.~\ref{fig:PDF_large}. The secondary peak arises from scattered photons partially compensating for LoS blockage during deep fading events, thereby raising the irradiance floor and concentrating level crossings around this threshold.}

\bl{Moreover, the turbid-water case consistently exhibits higher LCR values across the considered threshold range. This behavior interestingly contrasts with the small-bubble scenario, for which both water conditions yield comparable LCR values. The increased LCR under large-bubble conditions can be attributed to two concurrent effects.}
First, particle scattering is hypothesized to shorten bubble-induced fades by partially restoring the received irradiance during LoS blockage, thereby increasing the threshold crossing rate \bl{and decreasing} the scintillation index, as shown in Table~\ref{tab:pathloss}. This interpretation is supported by the AFD results presented in the next subsection. Second, the higher path loss in turbid water reduces the received SNR, potentially introducing additional noise-induced crossings, specially in low-threshold scenarios. This hypothesis, however, could not be verified experimentally because the transmitter laser was already operating at its maximum output power. \bl{Further measurements under matched-SNR conditions are required to distinguish between these two contributions.}\\

\subsubsection{Analytical LCR results}
Regarding the analytical results, \bl{some differences can be observed between both approaches presented in Section~\ref{sec:LCR_AFD} with respect to the experimental results. Approach 1 consistently overestimates the LCR relative to the measurements at low threshold values, whereas Approach 2, by explicitly accounting for the sampling period, provides a closer match to the experimental results over the entire threshold range.}

\begin{figure}[!t]
\centering
\begin{minipage}{1\columnwidth}
    \centering
    \includegraphics[width=\linewidth]{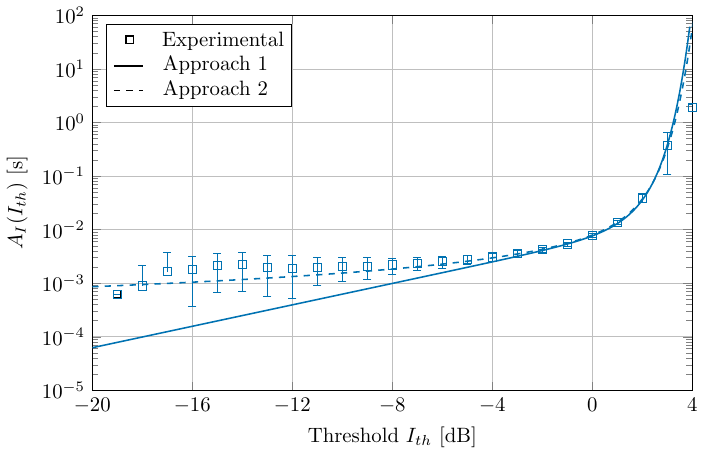}
    (a) 
\end{minipage}
\hfill
\begin{minipage}{1\columnwidth}
    \centering
    \includegraphics[width=\linewidth]{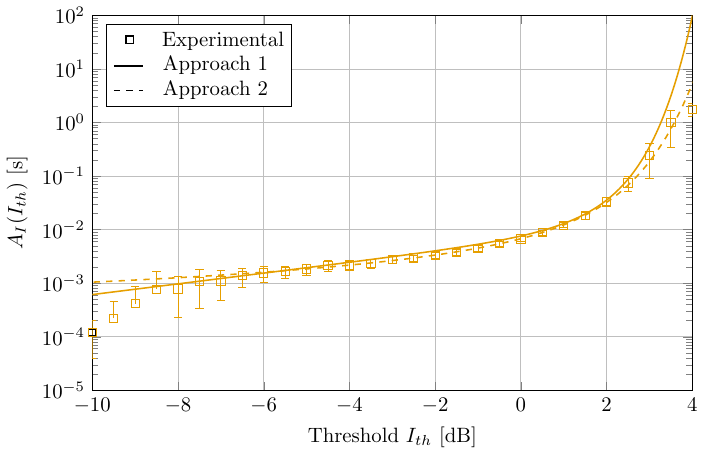}
    (b) 
\end{minipage}
\caption{Average AFD (in seconds) versus threshold $I_{\mathrm{th}}$ for (a) tap water and (b) turbid water ($200$~mg of antacid) in the presence of \textbf{small air bubbles}.}
\label{fig:AFD_small}
\end{figure}
\begin{figure}[!t]
\centering
\begin{minipage}{1\columnwidth}
    \centering
    \includegraphics[width=\linewidth]{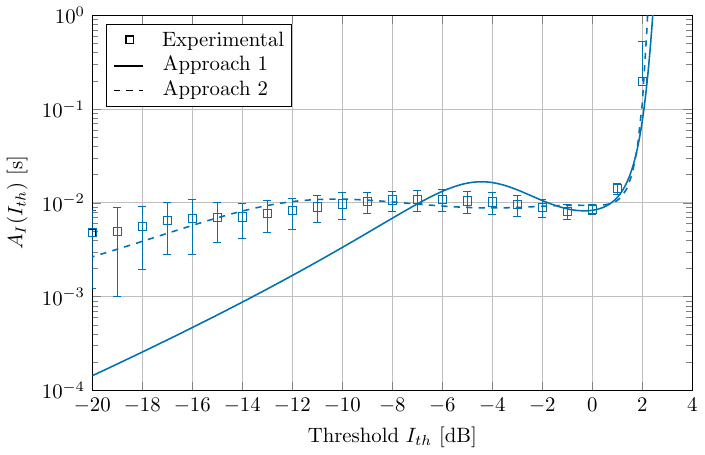}
    (a) 
\end{minipage}
\hfill
\begin{minipage}{1\columnwidth}
    \centering
    \includegraphics[width=\linewidth]{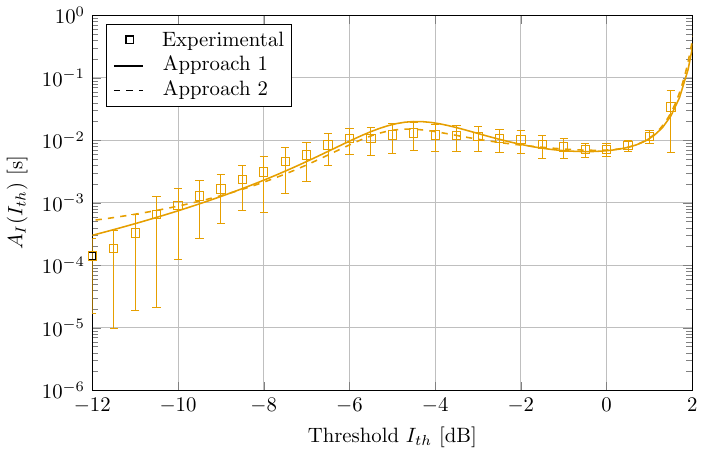}
    (b) 
\end{minipage}
\caption{Average AFD (in seconds) versus a threshold $I_{\mathrm{th}}$ for (a) tap water and (b) turbid water ($200$~mg of antacid) in the presence of \textbf{large air bubbles}.}
\label{fig:AFD_large}
\end{figure}

As discussed in \cite{ramirez-espinosa_extension_2018, lopez-martinez_average_2015}, the Rice approach neglects the finite sampling period and therefore counts all threshold crossings. Furthermore, under the experimental conditions considered here, this overestimation is more pronounced at low thresholds, where fades are shorter and missed crossings due to finite sampling become more frequent (as shown later in Figs.~\ref{fig:AFD_small} and~\ref{fig:AFD_large}). \bl{Additionally, the estimation of $\sigma^2_{\dot{I}}$ from the measured signal introduces a further source of overestimation in Approach~1. As shown in (\ref{eq:PSD}), the integrand is weighted by $\omega^2$, which increasingly amplifies the contribution of high-frequency components to the estimate of $\sigma^2_{\dot{I}}$. 
Since bubble-induced irradiance fluctuations are predominantly concentrated at low frequencies, the high-frequency region of the estimated PSD is mainly dominated by residual high-frequency noise. Consequently, the $\omega^2$ weighting disproportionately amplifies this noise contribution, resulting in an overestimation of $\sigma^2_{\dot{I}}$ and, consequently, of the LCR. Approach~1 is also affected by errors arising from the parametric fitting of the analytical PDF. These deviations are most pronounced at low thresholds, where rare deep-fade events are less accurately represented by the fitted PDF under the MSE fitting criterion. A more detailed analysis of this effect is provided in~\cite{fernandez_how_2025}.}

The discrepancies between Approach 2 and the measurements mainly arise from its dependence on the fitted parameters (requiring the fitted CDF and bivariate CDF). Estimation errors due to numerical fitting inaccuracies are amplified because, as $T_S\rightarrow 0$, both the numerator and denominator in (\ref{eq:LCR_bivariateApproach}) approach zero, making the LCR highly sensitive to small fitting errors. \bl{However, Approach~2 remains within the experimental variability bands across the entire threshold range, indicating good agreement between the model predictions and the measurements within the observed experimental variability.}

\subsection{Average Fade Duration Analysis}
Figures~\ref{fig:AFD_small} and \ref{fig:AFD_large} present the AFD for the small- and large-bubble scenarios, respectively. Since the AFD is inversely proportional to the LCR for a given threshold from (\ref{eq:AFD}), it exhibits the complementary trends discussed previously for LCR.
For small-bubbles, both analytical approaches accurately reproduce the quasi-exponential increase in the AFD with \mbox{$I_{\mathrm{th}}$}, while for large-bubbles they also capture the local minima induced by the bimodal irradiance distribution. Moreover, \bl{the AFD results in the large-bubble case} support the hypothesis that increased turbidity reduces the duration of bubble-induced fades through particle scattering \bl{when LoS blockage is not negligible}.
For example, at \mbox{$I_{\mathrm{th}} = -10 $~dB} in the large-bubble scenario, the AFD decreases from approximately $10$~ms in tap water to $1$~ms in turbid water.

Consistent with the LCR results, Approach~1 underestimates the AFD at low thresholds because it overestimates the crossing rate, whereas Approach~2 remains accurate over a wider threshold range.
Note, similar discrepancies between theoretical and experimental AFD have also been reported for RF shadowed fading channels using parameters estimated by nonlinear least-squares fitting~\cite{cotton_second-order_2016}.

\section{Packet Outage Probability Analysis}
To illustrate the usefulness of the derived second-order statistics, we evaluate the performance metric of packet-outage-probability (POP) of the UOWC link, which cannot be predicted from first-order statistics alone. Since the coherence time of air-bubble-induced fading is much longer than the symbol period at data rates above a few Mb/s, transmission errors occur in bursts spanning entire packets. Consequently, the average BER reported in \cite{salcedo-serrano_performance_2024} does not fully characterize the overall link performance.

\begin{figure}[!t]
\centering
\begin{minipage}{1\columnwidth}
    \centering
    \includegraphics[width=\linewidth]{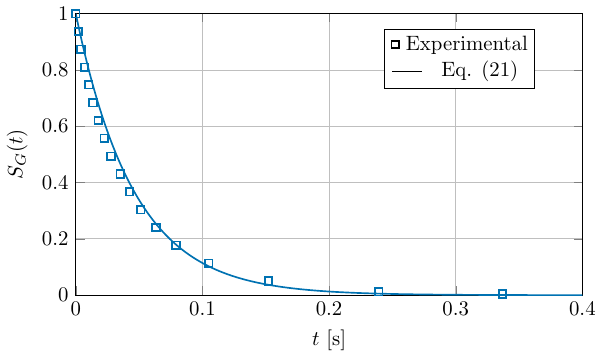}
    (a) 
\end{minipage}
\hfill
\begin{minipage}{1\columnwidth}
    \centering
    \includegraphics[width=\linewidth]{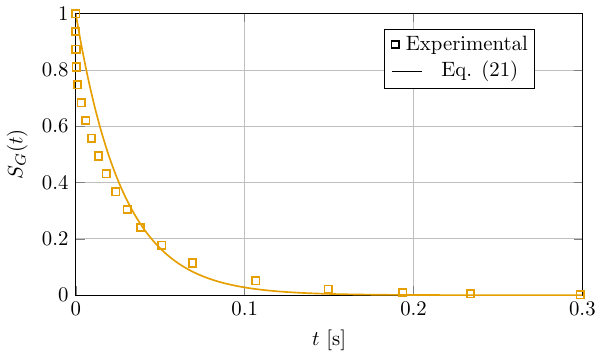}
    (b) 
\end{minipage}
\caption{\bl{Survival function for (a) tap water and (b) turbid water ($200$~mg of antacid) in the presence of \textbf{small air bubbles}.}}
\label{fig:PER_MarkovAssumption_smallBubbles}
\end{figure}

\begin{figure}[!t]
\centering
\begin{minipage}{1\columnwidth}
    \centering
    \includegraphics[width=\linewidth]{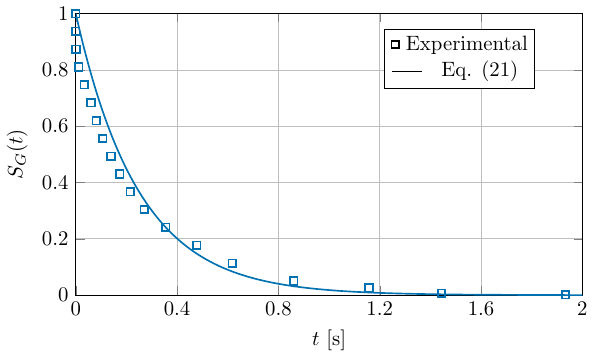}
    (a) 
\end{minipage}
\hfill
\begin{minipage}{1\columnwidth}
    \centering
    \includegraphics[width=\linewidth]{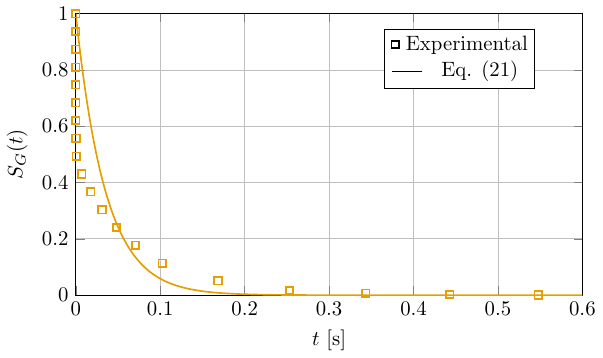}
    (b) 
\end{minipage}
\caption{\bl{Survival function for (a) tap water and (b) turbid water ($200$~mg of antacid) under \textbf{large-bubble} conditions.}}
\label{fig:PER_MarkovAssumption_largeBubbles}
\end{figure}

Following~\cite{fukawa_packet-error-rate_2012}, we model the channel as a two-state Markov process, where the channel is considered to be in the ``good'' state ($G$) when $I > I_{\text{th}}$ and in the ``bad'' state ($B$) otherwise. The threshold $I_{\text{th}}$ is defined as the minimum received irradiance required to maintain the instantaneous BER below the forward error correction (FEC) target of $3.8\times 10^{-3}$.
\bl{Hence, considering a threshold-based packet error probability}, a packet of duration $T_p$ is received successfully if the channel remains in the $G$ state throughout the entire packet. \bl{As a result}, the probability of successful packet reception, i.e., the complement of the POP, equals the probability that the channel starts in state $G$ and no $G\to B$ transition occurs during $T_p$. The stationary probability of state $G$ is given by 
\begin{equation}
    P_G = 1 - F_I(I_{\text{th}}).
    \label{eq:good}
\end{equation}
\bl{Assuming that the durations of the $G$-state periods follow an exponential distribution}, the corresponding survival probability over $T_p$ is \mbox{\cite[Eq.~(57)]{fukawa_packet-error-rate_2012}}
\begin{equation}
    S_{G} = \exp\left(-\frac{N_I(I_{\text{th}})\, T_p}{P_G}\right),
    \label{eq:g2g}
\end{equation}
where $N_I(I_{\text{th}})/P_G$ is the transition rate from $G$ to $B$ under the Markov assumption. \bl{To assess the validity of this assumption, the empirical survival function of the $G$-state durations was estimated directly from the experimental measurements. For a fixed threshold $I_{\text{th}}$, the irradiance record was converted into a binary sequence $X_n \in \{G, B\}$, as described above. All consecutive sojourns in state $G$ were then identified, and their durations were used to compute the empirical survival function $\widehat{S_{G}}(t)$. Figures \ref{fig:PER_MarkovAssumption_smallBubbles} and \ref{fig:PER_MarkovAssumption_largeBubbles} compare $\widehat{S_{G}}(t)$ with the exponential model in (\ref{eq:g2g}) for both tap- and turbid-water conditions under small- and large-bubble scenarios, respectively. As observed, the empirical survival functions generally follow the exponential prediction across the considered range of durations, supporting the memoryless assumption and, consequently, the use of a two-state continuous-time Markov model to characterize the $G$-state statistics under the considered experimental conditions. }

\bl{Therefore, combining the survival probability of the $G$ state with the probability of being in state $G$ yields}
\begin{equation}
    \mathrm{POP}\,(I_{\text{th}}, T_p)
    = 1 - P_G \exp\!\left(- \frac{N_I(I_{\text{th}})\, T_p}{P_G}\right).
    \label{eq:per}
\end{equation}

\bl{To assess the validity of (\ref{eq:per}), the $\mathrm{POP}$ is also estimated directly from the experimental measurements. Specifically, the empirical $\mathrm{POP}$ is computed as the fraction of sliding windows of duration $T_p$ containing at least one instance in which the irradiance falls below $I_{\mathrm{th}}$, as follows:
\begin{equation}
\begin{split}
        \widehat{\mathrm{POP}}(I_{\mathrm{th}}, T_p) & = 1 - \frac{1}{N - m + 1}  \\ & \times \sum_{n=1}^{N-m+1} 
        \mathbf{1}\left\{\min_{0 \leq k < m} I[n+k] > I_{\mathrm{th}}\right\},
        \label{eq:PER_emp}
    \end{split}
\end{equation}
where $m = \lfloor T_p / T_s \rfloor$ is the number of samples corresponding to a packet of  duration $T_p$ and $\mathbf{1}\{\cdot\}$ is the indicator function}.

\begin{figure}[!t]
\centering
\begin{minipage}{1\columnwidth}
    \centering
    \includegraphics[width=\linewidth]{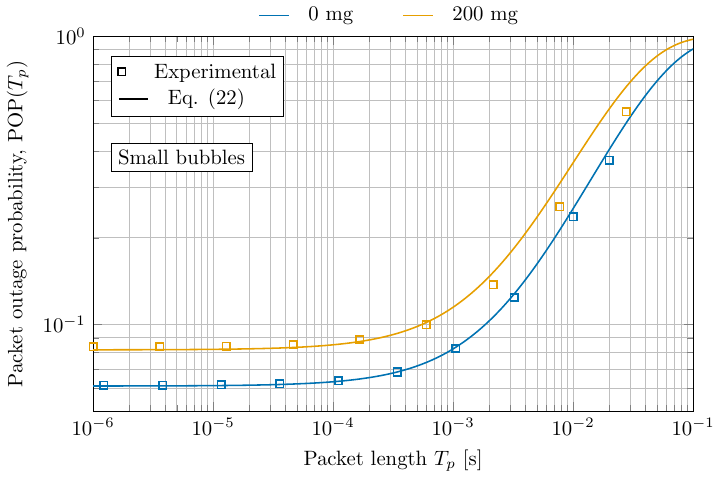}
    (a) 
\end{minipage}
\hfill
\begin{minipage}{1\columnwidth}
    \centering
    \includegraphics[width=\linewidth]{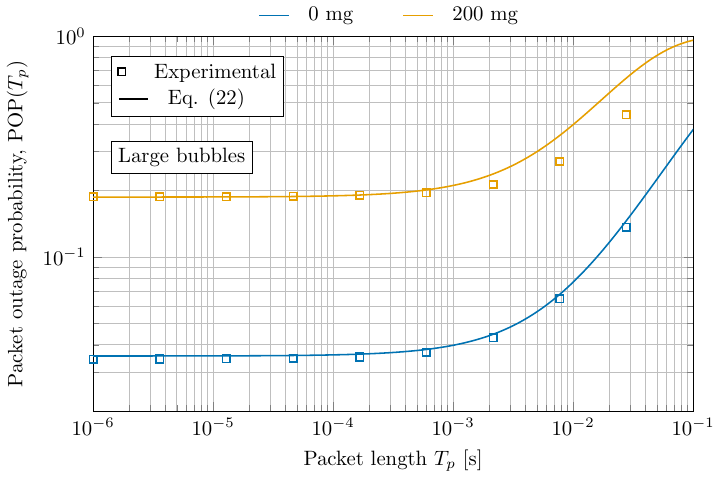}
    (b) 
\end{minipage}
\caption{\bl{Packet outage probability versus packet duration $T_p$ under (a) small- and (b) large-bubble conditions for tap water and turbid water ($0$ and $200$~mg of antacid, respectively).
}}
\label{fig:PER}
\end{figure}

Figure~\ref{fig:PER} presents the POP as a function of the packet duration $T_p$ for the small- and large-bubble scenarios under both water turbidity conditions using (\ref{eq:PER_emp}) and for the analytical Approach~2, which was shown to provide closer agreement with the experimental LCR across the entire threshold range. Also, Table~\ref{table:PER_summ} summarizes the corresponding SNR values, the decision threshold required to achieve the target BER of $3.8\times10^{-3}$, together with $N_I(I_{\text{th}})$ and $A_{I}(I_{\text{th}})$.

\begin{table}[ht]
    \centering
    \renewcommand{\arraystretch}{1.3}
    \setlength{\tabcolsep}{4pt}   
    \caption{SNR and intensity thresholds for $\text{BER} = 3.8\times10^{-3}$, level crossing rate and average fade duration for small- and large-bubble scenarios under tap and turbid water ($200$~mg antacid).}
    \label{table:PER_summ}
    \begin{tabular}{lcccc}
        \hline
        & \multicolumn{2}{c}{\textbf{Small bubbles}} 
        & \multicolumn{2}{c}{\textbf{Large bubbles}} \\
        \cline{2-3} \cline{4-5}
        & {0 mg} & {200 mg} & {0 mg} & {200 mg} \\
        \hline\hline
        \textbf{SNR$_{\text{th}}$} 
        & $13.99$ dB & $12.28$ dB & $33.34$ dB & $8.35$ dB \\
        \textbf{$I_{\text{th}}$} 
        & $-4.24$ dB & $-3.38$ dB & $-13.91$ dB & $-1.42$ dB \\
        \hline
        $N_I(I_{\text{th}})$
        & $21.1$ & $33.1$ 
        & $4.28$ & $24.5$ \\
        $A_{I}(I_{\text{th}})$
        & $2.84\times10^{-3}$ & $2.43\times10^{-3}$ 
        & $8.31\times10^{-3}$ & $7.68\times10^{-3}$ \\
        \hline
    \end{tabular}
\end{table}

As expected, the POP increases monotonically with $T_p$. In the limit of $T_p \to 0$, the POP reduces to the outage probability, $F_I(I_{\text{th}})$, since an infinitesimally short packet is lost only if it starts during a fade. As $T_p$ increases, the probability of a $G\to B$ transition during packet reception also increases, resulting in a higher POP. Consequently, channels with the same outage probability but different crossing rates exhibit different packet-level performances, highlighting again the importance of accounting for second-order statistics.

\bl{In the tap-water case (blue curves), the analytical POP closely matches the empirical results over the entire range of packet durations considered. Under turbid-water conditions (orange curves), good agreement is maintained up to $T_p \approx 2\times 10^{-3}$~s, beyond which the analytical model progressively overestimates the empirical POP. This slight discrepancy can be attributed to deviations of the $G$ -state survival function from the exponential distribution assumed by the Markov model under turbid-water conditions, for both small- and large-bubble scenarios. Nevertheless, the analytical expression accurately captures the overall dependence of the POP on packet duration and, importantly, reproduces the consistently higher POP observed in turbid water compared with tap water across the considered packet-duration range for both bubble scenarios. Furthermore, the analytical model correctly identifies the knee point, i.e., the maximum packet duration below which the POP remains approximately constant and close to the outage probability $F_I{(I_{\text{th}})}$, beyond which it increases markedly with $T_p$.} Lastly, although turbidity reduces the AFD, the lower received SNR increases the probability of G$\to$B transitions through additional noise-induced threshold crossings, resulting in a higher POP. This again highlights the importance of considering second-order statistics: despite the lower BER and SNR required to achieve the FEC limit as reported in~\cite{salcedo-serrano_performance_2024}, the higher transition rate degrades packet-level performance. Therefore, link-layer metrics such as POP provide a more comprehensive characterization of the channel than physical-layer metrics based solely on first-order statistics.


\section{Conclusions}
\label{sec:conclusions}
In this work, the LCR and AFD of bubble-impaired UOWC channels were characterized under different air bubble sizes and water turbidity conditions, using two analytical approaches that were both experimentally validated. \bl{For the considered fitting parameters, Approach~1, based on Rice's formula, tends to overestimate the LCR and, consequently, underestimate the AFD. This discrepancy is particularly pronounced at low threshold levels. In contrast,} Approach~2 \bl{defines the LCR for sampled fading channels in terms of the fading-envelope CDF and its bivariate CDF, thereby more explicitly accounting for sampling effects and achieving closer agreement for LCR and AFD with the experimental results.} \bl{Both approaches show that water turbidity has a negligible effect on the fade statistics under small-bubble conditions, whereas in large-bubble scenarios}, the turbid water scenarios exhibit higher LCR and lower AFD values than the corresponding tap water cases, i.e., shorter but more frequent deep fades. \bl{This behavior may result from the combined effects of particle-induced scattering, which mitigates LoS blockage through additional beam spreading, and the increased relative contribution of measurement noise associated with the lower received power. Further measurements under matched-SNR conditions are required to isolate the respective contributions of these two effects.}
The derived second-order statistics were then applied to evaluate the POP using a two-state Markov channel model. Despite the shorter fade durations observed in turbid water, the increased fade rate results in a higher POP than in tap water. These findings demonstrate that packet-level performance metrics can diverge significantly from bit-level metrics such as BER, underscoring the importance of second-order statistics for the design and evaluation of UOWC link-layer protocols.

Future work will extend the proposed two-state Markov model to higher-order channel models that more accurately capture the temporal dynamics of bubble-impaired UOWC channels. Such models could support the development of adaptive link-layer mechanisms, including coding and interleaving, coding and retransmission strategies, to mitigate the impact of bubble-induced fading.



\vfill

\end{document}